\documentclass[conference]{IEEEtran}

\usepackage{booktabs}
\usepackage{graphicx}
\usepackage{amsmath}
\usepackage{xspace}
\usepackage{url}

\newcommand{\tool}{RAGScope\xspace}
\newcommand{\rz}{RAGScope-Z\xspace}
\newcommand{\rc}{RAGScope-C\xspace}
\newcommand{\re}{RAGScope-E\xspace}
\graphicspath{{figures/}}

\begin{document}
\raggedbottom

\title{RAGScope: A Leakage-Controlled, Cost-Aware Evidence-Gating Protocol for RAG Hallucination Triage}

\author{
\IEEEauthorblockN{Zeming Liu\IEEEauthorrefmark{1}, Qibai Chen\IEEEauthorrefmark{2},
Jingtao Zhang\IEEEauthorrefmark{3}, and Hang Lyu\IEEEauthorrefmark{1}}
\IEEEauthorblockA{\IEEEauthorrefmark{1}Brown University, United States}
\IEEEauthorblockA{\IEEEauthorrefmark{2}Independent Researcher, United States}
\IEEEauthorblockA{\IEEEauthorrefmark{3}Georgia Institute of Technology, United States}
}

\maketitle

\begin{abstract}
Retrieval-augmented generation (RAG) systems need inexpensive ways to route
generated answers: accept low-risk outputs, review uncertain ones, and reserve
strong verifiers for the expensive tail. We present \tool, a leakage-controlled
protocol for evaluating local evidence gates that use only the task input,
retrieved context, and answer text. The protocol combines context-grouped
splits, fold-scoped preprocessing, group bootstrap intervals, deployment
operating points, end-to-end runtime, and explicit source-shift stress tests. On
three RAGTruth tasks, the enhanced gate \re reaches 0.798 AUROC and 0.660
average precision (AP) in pooled grouped cross-validation. Its pooled AP exceeds
ROUGE-L by 0.034 with a 95\% context-group interval of [0.002, 0.064], although
the AUROC gain is not significant and ROUGE-L remains stronger on
data-to-text. At a top-10\% review budget, \re attains 0.748 precision; accepting
the lowest-risk 50\% yields 0.141 residual unfaithfulness. \re runs in
6.22 ms/example on CPU, versus 145.75 and 223.07 ms/example for the tested
DeBERTa-NLI and HHEM settings. A 14,900-example HaluBench stress test exposes
the deployment boundary: an in-domain calibrated gate reaches 0.879 AUROC, but
leave-source-out calibration averages only 0.466. Target-only calibration
recovers to 0.675 AUROC with 100 labels per source and 0.685 with 200. Cheap evidence gates are
therefore useful routing components, but learned calibration must be validated
and adapted within the target domain.
\end{abstract}

\section{Introduction}

RAG systems are often evaluated after an answer has already been produced:
a verifier, judge, or human reviewer decides whether the answer is supported by
retrieved evidence. In practice, however, teams also need a routing decision.
If every answer is sent to a large verifier or human reviewer, evaluation is
expensive and slow; if every answer is accepted, unsupported content can pass
silently. The operational question is therefore selective: which outputs are
safe enough to accept locally, and which should be escalated?

Existing factuality and RAG-evaluation methods include NLI-style consistency
models, sampling-based detectors, RAG-specific metric suites, and LLM judges
\cite{kryscinski-etal-2020-evaluating,laban-etal-2022-summac,manakul-etal-2023-selfcheckgpt,es-etal-2024-ragas,zheng2023judgingllmasajudge}.
These methods are important, but they can add model-serving dependencies,
latency, and cost to every development loop. Lightweight lexical signals are
less expressive, but they are easy to run on every candidate answer and can be
used as a first-stage router before stronger verification.

The risk is that simple gates are easy to overstate. If the same context appears
in both train and test folds, if TF-IDF statistics are fit on the full dataset,
or if only pooled benchmark metrics are reported, a cheap gate can look more
general than it is. This paper therefore treats protocol design as part of the
contribution. We ask what a local RAG triage study should report so that its
claims remain useful under grouped examples, task heterogeneity, and realistic
deployment operating points.

We make four contributions:
\begin{itemize}
  \item We define \tool, a lightweight evidence-gating family together with a
  leakage-controlled evaluation protocol: grouped splits by source context,
  fold-scoped TF-IDF preprocessing, group bootstrap intervals,
  deployment-style review/accept operating points, and source-shift checks.
  \item We evaluate zero-shot coverage, ROUGE-L, TF-IDF, learned lexical gates,
  and an enhanced local gate (\re) on RAGTruth QA, summarization, and
  data-to-text outputs. The strongest supported gain is pooled AP and QA
  triage; \re does not dominate ROUGE-L on every task.
  \item We compare \re with local DeBERTa-NLI and HHEM verifier baselines,
  reporting accuracy and CPU runtime to expose verifier mismatch, cost, and
  task-level deployment boundaries.
  \item We stress-test six HaluBench sources. Cross-source transfer can fail
  catastrophically; target-only calibration with small labeled samples
  restores useful ranking performance.
\end{itemize}

\section{Evidence-Gating Protocol}

Let $q$ be the task input, $c$ the retrieved or provided context, and $a$ the
candidate answer. A gate returns a risk score $s(q,c,a)$ where larger values
mean higher probability of unfaithfulness. The score is used for routing rather
than final factual proof: high-risk examples can be reviewed or sent to a
stronger verifier, and low-risk examples can be accepted only at a chosen risk
tolerance.

\begin{figure*}[t]
\centering
\includegraphics[width=.96\textwidth]{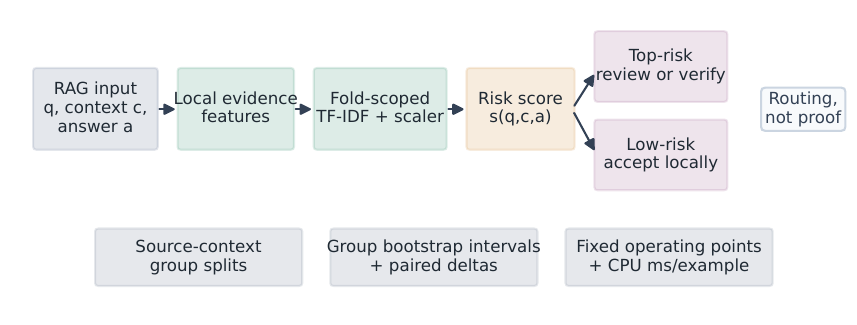}
\caption{\tool evaluates local evidence signals as a first-stage routing
protocol. The key controls are fold-scoped feature construction, context-grouped
evaluation, uncertainty intervals over groups, fixed review/accept operating
points, and runtime accounting.}
\label{fig:protocol}
\end{figure*}

Figure~\ref{fig:protocol} summarizes the intended use. The gate sits between a
RAG generator and an expensive verifier or reviewer. It is deliberately allowed
to be imperfect because it does not make the final factuality decision alone:
instead, it changes which examples consume expensive verification budget. This
routing view affects the evaluation. A model that slightly improves pooled AUROC
may still be unhelpful if it does not enrich the reviewed queue, and a model
with good review precision may still be unsafe if the automatically accepted
region has high residual risk.

\subsection{Routing Metrics}

Let $D=\{(q_i,c_i,a_i,y_i)\}_{i=1}^n$ and let $y_i=1$ denote an unfaithful
answer. For a review budget $\rho$, the policy sorts examples by decreasing
risk and reviews the top $\lceil \rho n\rceil$ examples. We report review
precision,
\[
\mathrm{Prec}_{\rho} =
\frac{\sum_{i\in \mathrm{Top}_{\rho}(s)} y_i}
{|\mathrm{Top}_{\rho}(s)|},
\]
which asks how concentrated the flagged queue is. For an automatic-acceptance
budget $\alpha$, the policy sorts by increasing risk and accepts the lowest-risk
$\lceil \alpha n\rceil$ examples. We report accepted risk,
\[
\mathrm{Risk}_{\alpha} =
\frac{\sum_{i\in \mathrm{Low}_{\alpha}(s)} y_i}
{|\mathrm{Low}_{\alpha}(s)|}.
\]
These two quantities are not substitutes for AUROC or AP; they expose the
threshold behavior that a deployment team must choose before using a gate.

\subsection{Lightweight Evidence Features}

\rz uses monotone evidence-missing signals. The simplest baseline is the
fraction of non-stopword answer tokens that do not occur in the context:
\[
s_{\mathrm{uncov}}(a,c)=
1-\frac{|T(a)\cap T(c)|}{|T(a)|}.
\]
The base feature family also includes answer length, context length, answer
coverage, rare-token coverage for tokens of length at least six, answer-context
Jaccard overlap, question-answer Jaccard overlap, longest contiguous supported
run, best sentence-level support, numeric coverage, and numeric count. Tokens
are lowercased alphanumeric spans; sentence boundaries use punctuation and line
breaks; numeric matching strips commas.

\rc trains a logistic regression model over the base features:
\[
s_{\mathrm{cal}}(q,c,a)=\sigma(w^\top\phi(q,c,a)+b).
\]
The model uses standardization, balanced class weights, $L_2$ regularization,
and a fixed random seed. It is intended for settings with local labels and a
fixed operating policy.

\subsection{Enhanced Local Gate}

\re adds four fully local support features to the base family: ROUGE-L risk,
token-F1 risk, TF-IDF answer-context risk, and maximum sentence-level TF-IDF
risk. ROUGE-L uses longest common subsequence recall over the same non-stopword
tokens; token-F1 uses bag overlap; TF-IDF uses scikit-learn English stop words,
L2-normalized unigram/bigram vectors, cosine similarity, and \texttt{min\_df=1}.
No remote service, LLM, or NLI model is
called. Because TF-IDF statistics are data dependent, \re is evaluated with an
outer grouped split: in each fold, TF-IDF vocabularies and IDF weights are fit
only on the training fold and are then used to transform held-out examples.
Standardization and logistic parameters are also fit inside the training fold
using scikit-learn logistic regression with \texttt{lbfgs}, $C=1$, balanced
class weights, 1,000 maximum iterations, and random seed 13.

\subsection{Protocol Requirements}

We evaluate gates with the checklist in Table~\ref{tab:protocol}. First, all
learned RAGTruth models use
StratifiedGroupKFold with group key \texttt{task:source\_info}, so the six
answers attached to the same source context never cross train/test boundaries.
Second, paired bootstrap intervals resample source-context groups rather than
individual answers. Third, operating points are specified by workload fractions
before evaluation: review the top-risk 5--30\% or accept the lowest-risk
50--90\%. Fourth, cost-quality comparisons report wall-clock CPU runtime per
example for complete local inference, not only model scoring. Fifth, a learned
gate is not treated as source-general merely because random or grouped
cross-validation is strong: calibration is also evaluated by holding out entire
dataset sources and by measuring recovery from small target-labeled samples.

\begin{table}[!htbp]
\centering
\caption{\tool protocol checklist. Each item is implemented by the released
scripts and reported in the artifact manifest.}
\label{tab:protocol}
\scriptsize
\begin{tabular}{@{}ll@{}}
\toprule
Risk & Required control \\
\midrule
Context leakage & Group splits by source context \\
Preprocess leakage & Fit TF-IDF/scalers inside folds \\
Uncertain gains & Group bootstrap and paired deltas \\
Deployment mismatch & Report review/accept points \\
Verifier cost & Measure end-to-end ms/example \\
Source shift & Hold out sources; vary target labels \\
\bottomrule
\end{tabular}
\end{table}

\section{Experimental Design}

\textbf{RQ1.} How far do cheap local evidence signals go on RAGTruth triage?

\textbf{RQ2.} Under leakage-controlled grouped evaluation, where does \re
improve over uncovered-token fraction and ROUGE-L, and where does it fail?

\textbf{RQ3.} What review and automatic-acceptance operating points does the
gate enable?

\textbf{RQ4.} How does the local gate compare with tested local verifier
settings in accuracy and runtime?

\textbf{RQ5.} How does learned evidence calibration behave under source shift,
and how many target-domain labels are needed to recover?

\subsection{Datasets}

RAGTruth provides QA, summarization, and data-to-text outputs with
response-level faithfulness labels derived from annotated hallucination spans
\cite{niu-etal-2024-ragtruth}. We evaluate the processed public test splits:
900 examples per task and 2,700 examples total. Positive labels indicate
unfaithful or hallucinated outputs.

HaluBench contains 14,900 context--question--answer examples from six source
datasets spanning reading comprehension, finance, biomedical QA, and
hallucination benchmarks \cite{ravi2024lynx}. Its source identifiers make it
useful for a deliberately difficult transfer test. We first report five-fold
in-domain CV for \rc, the same standardized logistic feature gate without
TF-IDF features. We then hold out each source, train on the other five, and
evaluate both the transferred calibrated gate and the training-free
uncovered-token score. Finally, for each target source and
$k\in\{25,50,100,200\}$, we reserve a fixed stratified 20\% target test set,
draw a prevalence-preserving stratified labeled sample from the remaining
80\%, fit \rc only on that sample, and repeat over ten seeds.

\subsection{Baselines and Metrics}

Cheap local baselines include uncovered-token fraction, token-F1 risk, ROUGE-L
risk, TF-IDF answer-context risk, and maximum sentence-level TF-IDF risk. Learned
baselines include a token-coverage logistic model, a base all-feature logistic
model, and \re. Stronger local verifier baselines include a DeBERTa-v3-small NLI
cross-encoder \cite{he2021debertav3} and Vectara HHEM
\cite{vectara-hhem}. Both are run locally on CPU. DeBERTa uses the top three
context sentences selected by lexical overlap with the answer. HHEM is reported
in a fuller setting using the complete context truncated to 512 tokens; a
top-three-sentence HHEM run is retained as an artifact-only lightweight
ablation.

We report AUROC, AP, context-group bootstrap 95\% intervals, paired deltas, and
operating points. AP is important because positive rates differ substantially
by task: 0.178 for QA, 0.227 for summarization, and 0.643 for data-to-text.
All random seeds are fixed in the released scripts. The artifact package
contains per-example scores, grouped bootstrap outputs, operating-point tables,
runtime JSON files, and the exact local model names used for verifier baselines.
This matters because the main result depends on out-of-fold scores rather than
on a single model fit to the full benchmark.

For learned local gates, every reported RAGTruth score is an out-of-fold score.
This choice makes operating-point curves meaningful: a reviewed example is never
scored by a model whose preprocessing or logistic parameters were fitted using
that example's source context. Bootstrap intervals use 1,000 resamples of
context groups for the combined setting and per-task groups for task-specific
metrics. Runtime is measured as complete local inference time per example,
including feature extraction and model scoring for cheap gates and full CPU
forward passes for verifier baselines.

For HaluBench leave-source-out results, 95\% intervals use 1,000 within-source
bootstrap resamples. Target-adaptation tables report macro averages over the
six sources and standard errors over ten repeated stratified target splits. This
stress test is intentionally stricter than RAGTruth grouped CV: it changes the
benchmark source itself rather than only withholding contexts.

\begin{table}[!htbp]
\centering
\caption{Experiment matrix. Learned gates use context-grouped splits; TF-IDF
features are fit inside each training fold.}
\label{tab:matrix}
\scriptsize
\resizebox{\linewidth}{!}{\begin{tabular}{@{}llll@{}}
\toprule
RQ & Dataset & Control & Metric \\
\midrule
RQ1 & RAGTruth tasks & cheap signals & AUROC, AP \\
RQ2 & RAGTruth all & group CV & AUROC, AP, delta \\
RQ3 & RAGTruth & risk cutoff & precision, risk \\
RQ4 & RAGTruth & verifier baselines & cost, accuracy \\
RQ5 & HaluBench & source holdout/adapt. & AUROC, AP \\
\bottomrule
\end{tabular}}
\end{table}

\section{Results}

\subsection{RQ1--RQ2: RAGTruth Triage}

\begin{table*}[t]
\centering
\caption{RAGTruth grouped results. Positive labels are unfaithful outputs. The
pooled setting is useful for a shared triage queue; macro averages expose
task-level generality.}
\label{tab:ragtruth}
\scriptsize
\begin{tabular}{lrrrrrr}
\toprule
Scope & Uncov. AUROC & Uncov. AP & ROUGE-L AUROC & ROUGE-L AP & \re AUROC & \re AP \\
\midrule
QA & 0.702 & 0.268 & 0.739 & 0.307 & 0.771 & 0.404 \\
Summarization & 0.676 & 0.391 & 0.694 & 0.402 & 0.682 & 0.377 \\
Data-to-text & 0.687 & 0.779 & 0.713 & 0.805 & 0.663 & 0.771 \\
Macro average & 0.689 & 0.479 & 0.715 & 0.505 & 0.706 & 0.517 \\
Combined & 0.779 & 0.594 & 0.792 & 0.624 & 0.798 & 0.660 \\
\bottomrule
\end{tabular}
\end{table*}

\begin{figure*}[t]
\centering
\includegraphics[width=.96\textwidth]{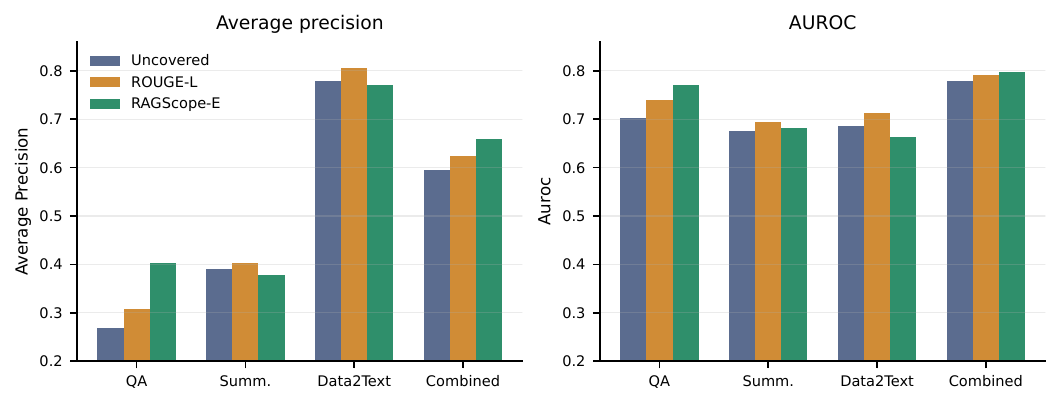}
\caption{RAGTruth grouped performance by task and pooled queue. \re improves
the pooled AP used for a shared triage queue and gives its clearest task-level
gain on QA, while ROUGE-L remains stronger on data-to-text.}
\label{fig:performance}
\end{figure*}

Table~\ref{tab:ragtruth} shows the main pattern. Uncovered-token fraction and
ROUGE-L are strong baselines. Figure~\ref{fig:performance} makes the same
pattern visible across both AP and AUROC. \re improves pooled AP to 0.660 and
improves QA substantially, but it is worse than ROUGE-L on data-to-text and does
not improve summarization AP. This makes the pooled claim useful but narrow:
\re is best viewed as a shared-queue triage gate, not as a task-universal
detector. The data-to-text result is especially important for claim discipline:
copy-heavy or schema-like outputs can reward long common subsequences, making a
simple ROUGE-L risk score hard to beat.

\begin{table}[!htbp]
\centering
\caption{Paired context-group bootstrap deltas for \re. Intervals are 95\%;
positive deltas favor \re.}
\label{tab:bootstrap}
\scriptsize
\begin{tabular}{llrr}
\toprule
Scope/base & Metric & Delta & 95\% CI \\
\midrule
Combined/uncov. & AUROC & 0.019 & [0.003, 0.034] \\
Combined/uncov. & AP & 0.064 & [0.029, 0.099] \\
Combined/ROUGE & AUROC & 0.006 & [-0.006, 0.017] \\
Combined/ROUGE & AP & 0.034 & [0.002, 0.064] \\
QA/ROUGE & AUROC & 0.031 & [0.006, 0.056] \\
QA/ROUGE & AP & 0.097 & [0.047, 0.148] \\
Data/ROUGE & AUROC & -0.050 & [-0.073, -0.027] \\
Data/ROUGE & AP & -0.035 & [-0.060, -0.008] \\
Macro/ROUGE & AUROC & -0.010 & [-0.024, 0.005] \\
Macro/ROUGE & AP & 0.013 & [-0.011, 0.036] \\
\bottomrule
\end{tabular}
\end{table}

\begin{figure}[!htbp]
\centering
\includegraphics[width=\linewidth]{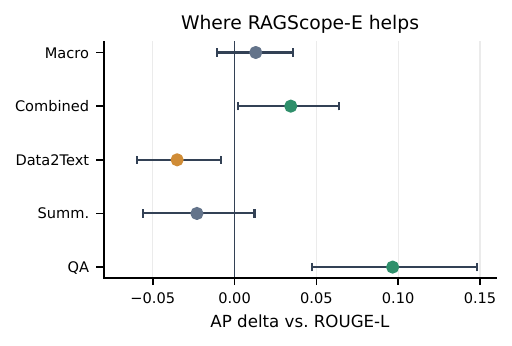}
\caption{Average-precision deltas for \re versus ROUGE-L with 95\%
context-group bootstrap intervals. The figure highlights the asymmetric result:
QA and pooled AP improve, while data-to-text favors ROUGE-L.}
\label{fig:bootstrap}
\end{figure}

Table~\ref{tab:bootstrap} clarifies statistical support. \re significantly
improves pooled AP over ROUGE-L and both pooled metrics over uncovered-token
fraction. Its pooled AUROC gain over ROUGE-L is not significant, and macro
deltas over ROUGE-L are not significant. Figure~\ref{fig:bootstrap} shows why a
single aggregate number would be misleading: the strongest positive AP delta is
in QA, while data-to-text has a negative interval. \textit{Takeaway:} the evidence
supports a cost-aware pooled triage use case and a strong QA result, not
per-task dominance.

\begin{table}[!htbp]
\centering
\caption{Grouped-CV ablation on RAGTruth combined.}
\label{tab:ablation}
\small
\begin{tabular}{lrr}
\toprule
Variant & AUROC & AP \\
\midrule
Token-coverage logit & 0.782 & 0.606 \\
Base all-feature logit & 0.792 & 0.636 \\
\re enhanced logit & 0.798 & 0.660 \\
\bottomrule
\end{tabular}
\end{table}

Table~\ref{tab:ablation} shows that the base lexical feature family already
captures much of the signal. The enhanced features add AP, but the gain is
incremental. This is why we frame the contribution as a protocol and operating
study rather than as a new neural verifier.

\subsection{RQ3: Operating Points}

\begin{figure*}[t]
\centering
\includegraphics[width=.96\textwidth]{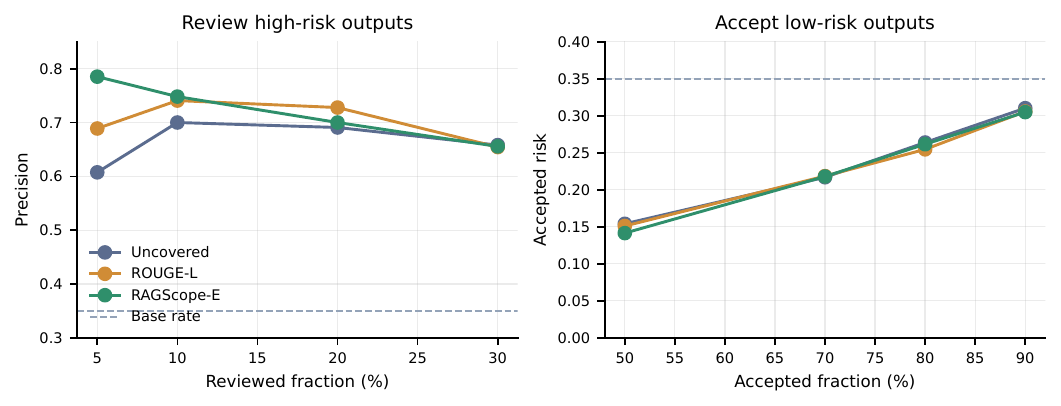}
\caption{Combined RAGTruth operating curves. Left: precision among examples
sent to review as the review budget grows. Right: residual unfaithfulness rate
among examples accepted locally as the acceptance budget grows. The dashed line
marks the combined base positive rate.}
\label{fig:operating}
\end{figure*}

\begin{table}[!htbp]
\centering
\caption{Operating points with context-group bootstrap intervals. Review is
top-risk 10\%; accept is lowest-risk 50\%.}
\label{tab:ops}
\scriptsize
\begin{tabular}{llcc}
\toprule
Scope & Model & Review precision & Accepted risk \\
\midrule
Combined & \re & .748 [.685,.807] & .141 [.119,.169] \\
Combined & ROUGE & .741 [.678,.807] & .151 [.127,.171] \\
QA & \re & .467 [.333,.600] & .067 [.044,.089] \\
QA & ROUGE & .344 [.200,.467] & .067 [.042,.093] \\
Summ. & \re & .456 [.344,.544] & .136 [.102,.171] \\
Summ. & ROUGE & .456 [.378,.556] & .124 [.093,.156] \\
Data & \re & .844 [.767,.911] & .536 [.493,.573] \\
Data & ROUGE & .856 [.778,.933] & .489 [.444,.529] \\
\bottomrule
\end{tabular}
\end{table}

Table~\ref{tab:ops} shows why operating points must be reported by task. The
combined top-10\% precision difference between \re and ROUGE-L is small and its
interval overlaps. Figure~\ref{fig:operating} adds the full combined curve:
review precision remains well above the 0.349 base rate for all three cheap
scores, while accepted risk rises as more examples are accepted automatically.
QA shows a larger review-routing gain for \re, whereas data-to-text is better
served by ROUGE-L, especially for automatic acceptance. \textit{Takeaway:}
\re is useful when the deployment has a shared high-risk review queue, but the
threshold policy should still be set per task when task identity is available.

\subsection{RQ4: Cost Versus Local Verifiers}

\begin{figure}[!htbp]
\centering
\includegraphics[width=\linewidth]{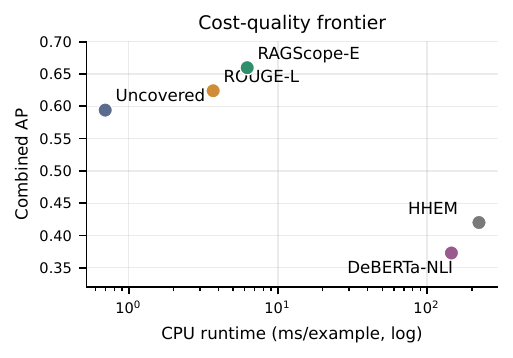}
\caption{Combined AP versus measured local CPU runtime. In this experiment,
\re lies on the cheap high-AP corner of the tested local methods, while the two
model-based verifier settings are slower and less accurate on RAGTruth.}
\label{fig:cost}
\end{figure}

\begin{table}[!htbp]
\centering
\caption{Accuracy-cost comparison on RAGTruth combined. Runtime is local CPU
milliseconds per example.}
\label{tab:cost}
\scriptsize
\begin{tabular}{lrrr}
\toprule
Model & AUROC & AP & ms/ex \\
\midrule
Uncovered fraction & 0.779 & 0.594 & 0.67 \\
ROUGE-L & 0.792 & 0.624 & 3.64 \\
\re & 0.798 & 0.660 & 6.22 \\
DeBERTa-NLI top-3 & 0.560 & 0.373 & 145.75 \\
HHEM full ctx. & 0.655 & 0.420 & 223.07 \\
\bottomrule
\end{tabular}
\end{table}

Table~\ref{tab:cost} compares local verifier baselines. HHEM improves when run
with full context truncated to 512 tokens rather than top-three evidence
sentences, but its context-group intervals remain below \re on the combined set:
0.655 [0.625, 0.687] AUROC and 0.420 [0.388, 0.457] AP. The result should not
be read as a universal rejection of NLI or HHEM; stronger prompting, answer
decomposition, or task-specific calibration may improve them. It does show that
off-the-shelf local verifiers are not automatically better on RAGTruth and are
substantially slower in a CPU-only evaluation loop. Figure~\ref{fig:cost}
visualizes the same tradeoff on a log runtime axis. The practical implication is
not that lexical gates replace verifiers, but that a low-millisecond gate can
screen every candidate before a verifier is invoked on the expensive tail.

The verifier result also illustrates why the evidence-selection policy should be
reported with the model name. The DeBERTa baseline receives only the top three
lexically selected context sentences, which keeps its input compact but can miss
supporting evidence or contradictions outside that subset. The HHEM run receives
a fuller context truncated to 512 tokens, which is more faithful to a RAG
setting but slower. \re does not solve these verifier-design choices; it offers
a cheap queueing layer whose runtime is small enough to run before making them.

\subsection{RQ5: Source Shift and Target Adaptation}

Random in-domain evaluation makes calibration look highly transferable.
On all 14,900 HaluBench examples, five-fold CV gives \rc 0.879 AUROC and
0.873 AP, compared with 0.761 and 0.639 for the training-free uncovered-token
score. Table~\ref{tab:shift} changes only the split: each source is now held out
in full.

\begin{table*}[t]
\centering
\caption{HaluBench source-shift stress test and target-label recovery.
Leave-source-out entries are metric [95\% bootstrap interval]. Adaptation
entries are macro mean $\pm$ standard error over ten target-only splits.}
\label{tab:shift}
\scriptsize
\begin{minipage}[t]{0.67\textwidth}
\centering
\textbf{(a) Leave one source out}\\[2pt]
\resizebox{\linewidth}{!}{%
\begin{tabular}{lrrrr}
\toprule
Held-out source & Zero AUROC & Zero AP & Cross-source AUROC & Cross-source AP \\
\midrule
DROP & .509 [.476,.544] & .497 [.461,.534] & .509 [.475,.544] & .505 [.463,.550] \\
FinanceBench & .487 [.451,.520] & .487 [.449,.526] & .494 [.459,.529] & .507 [.464,.556] \\
RAGTruth & .702 [.662,.741] & .268 [.229,.318] & .218 [.178,.255] & .110 [.095,.128] \\
covidQA & .731 [.709,.754] & .731 [.702,.760] & .594 [.558,.631] & .553 [.511,.603] \\
HaluEval & .894 [.886,.901] & .795 [.782,.809] & .492 [.481,.503] & .597 [.584,.609] \\
PubMedQA & .540 [.504,.576] & .536 [.491,.583] & .487 [.450,.520] & .490 [.451,.537] \\
\bottomrule
\end{tabular}}
\end{minipage}
\hfill
\begin{minipage}[t]{0.30\textwidth}
\centering
\textbf{(b) Macro target adaptation}\\[2pt]
\resizebox{\linewidth}{!}{%
\begin{tabular}{lrrr}
\toprule
Training & Labels & AUROC & AP \\
\midrule
zero-shot & 0 & .644 & .552 \\
cross-source & 0 & .466 & .460 \\
target-only & 25 & .639$\pm$.007 & .588$\pm$.008 \\
target-only & 50 & .665$\pm$.005 & .619$\pm$.005 \\
target-only & 100 & .675$\pm$.005 & .629$\pm$.006 \\
target-only & 200 & .685$\pm$.004 & .649$\pm$.006 \\
\bottomrule
\end{tabular}}
\end{minipage}
\end{table*}

The pooled in-domain result does not survive source transfer. Cross-source
calibration averages only 0.466 AUROC and 0.460 AP, below the zero-shot macro
values of 0.644 and 0.552. The failure is not limited to a mildly harder source:
the calibrated ranking inverts on held-out RAGTruth (0.218 AUROC) and falls near
random on HaluEval, even though uncovered-token risk remains informative on both
sources. This pattern indicates that the multifeature calibration learned
source-specific relationships among answer format, coverage, and labels rather
than a source-invariant hallucination boundary.

Small target-labeled samples provide a practical recovery path. With 25 labels
per source, target-only calibration improves macro AP to 0.588, although its
0.639 AUROC remains near the 0.644 zero-shot value. At 50 labels it surpasses
zero-shot AUROC, reaching 0.665 AUROC and 0.619 AP; 200 labels raise these to
0.685 and 0.649. The operational rule is therefore asymmetric: use monotone
zero-shot signals when no target labels exist, and deploy a learned gate only
after target-scoped fitting, grouped validation, and threshold selection.
Source shift is not a secondary limitation of \tool; it is one of the
protocol's required checks.

\section{Discussion and Limitations}

\subsection{Deployment Guidance}

The safest way to use \tool is to treat it as a routing contract rather than a
standalone detector. A team first chooses a budgeted policy: for example, review
the top 10\% highest-risk outputs, or accept only the lowest-risk 50\% and send
the rest to a stronger verifier. The policy is then validated on grouped
examples using the same preprocessing boundary that will be used in deployment.
If task identity is available, thresholds should be selected per task because
the QA, summarization, and data-to-text results have different base rates and
different best cheap baselines. If the learned gate was calibrated on another
source, it should remain disabled until a target-labeled audit shows that its
ranking direction and operating points transfer.

The protocol is intentionally compatible with stronger downstream checks. A
system can run \re on every answer, send the riskiest outputs to a verifier or
human reviewer, and periodically audit a sample of automatically accepted
outputs to recalibrate the acceptance budget. In this setting, the most useful
summary is not a single AUROC number. It is a small operating report: review
precision at the chosen workload, residual accepted risk, confidence intervals
over context groups, milliseconds per example, and a leave-source-out or
target-label stress test.

\subsection{Limitations}

\tool is intentionally shallow. It cannot prove factuality, solve multi-hop
reasoning, or catch contradictions that reuse the same evidence tokens. It is
also sensitive to the operational queue: pooled metrics can improve when tasks
have different base rates, while per-task metrics reveal where a gate is weak.
For this reason, the paper reports both combined and macro/task-level results.

The verifier comparison is also scoped. We used local CPU inference,
top-three lexical evidence for DeBERTa-NLI, and full-context truncation for
HHEM. A production verifier could use better retrieval, answer decomposition,
longer context windows, or calibration. Our claim is only that these
off-the-shelf local verifiers did not dominate cheap gates under the tested
local protocols.

The main RAGTruth experiments use public benchmark test splits with grouped
cross-validation rather than a final benchmark hidden from feature and model
selection. HaluBench adds source-level stress but is itself a public benchmark,
and its target-adaptation study samples labels from the same source that is
later evaluated. The study therefore measures source adaptation, not temporal
drift or production generalization. Future work should repeat the full operating
report on time-separated and product-specific logs.

\section{Related Work}

RAGTruth provides response-level and span-level hallucination annotations for
QA, summarization, and data-to-text generation
\cite{niu-etal-2024-ragtruth}; HaluEval broadens hallucination evaluation
across generated question-answer examples \cite{li-etal-2023-halueval}.
HaluBench was introduced with Lynx to evaluate hallucination judges across
heterogeneous sources \cite{ravi2024lynx}; cross-domain transfer is also an
explicit robustness criterion in other detector settings, including
cross-species ultrasonic-vocalization detection \cite{wei2026usvexplorer}.
RAGAS evaluates RAG pipelines with
metrics for faithfulness, answer relevance, and context quality
\cite{es-etal-2024-ragas}. \tool differs by studying a low-cost routing score
and by making grouped validation, operating points, cost, and source transfer
joint requirements of the claim.

Factuality and hallucination detection methods often use model-based verifiers.
NLI-style approaches have been effective for summarization inconsistency
detection \cite{kryscinski-etal-2020-evaluating,laban-etal-2022-summac}, and
TRUE re-evaluates factual consistency metrics across tasks
\cite{honovich-etal-2022-true-evaluating}. SelfCheckGPT detects hallucinations
through black-box sampling consistency \cite{manakul-etal-2023-selfcheckgpt},
while FActScore decomposes long-form generations into atomic facts
\cite{min-etal-2023-factscore}. LLM-as-judge methods can provide flexible
evaluation but introduce cost and judge-bias concerns
\cite{zheng2023judgingllmasajudge}. \tool is complementary: it gives a
millisecond-scale signal before invoking stronger judges.

Selective classification studies models that trade coverage for lower error by
deferring uncertain examples \cite{geifman2017selectiveclassification}. Our
operating points adapt this perspective to RAG evaluation: the system can review
high-risk outputs or automatically accept only low-risk outputs.

\section{Conclusion}

This paper presents \tool, a leakage-controlled, cost-aware protocol for local
RAG hallucination triage. Under grouped RAGTruth evaluation, \re improves pooled
AP and QA triage, but not every task, and provides a low-millisecond routing
signal before the tested local verifiers. HaluBench then exposes the more
important boundary: strong in-domain calibration can invert under source
transfer, while 50--200 target labels recover useful ranking performance.
Cheap evidence gates are therefore viable first-stage routers only when their
evaluation reports task-specific operating points, runtime, uncertainty, and
target-domain transfer rather than a single pooled benchmark score.

\IEEEtriggeratref{12}
\bibliographystyle{IEEEtran}
\bibliography{references}

\end{document}